\documentstyle[12pt,epsf]{article}
\newcommand{\Li}{\textrm {Li}}
\newcommand{\ams}{\alpha _{\overline {MS}}}
\newcommand{\as}{\alpha _s}

\begin{document}

\thispagestyle{empty}
\begin{flushright}
MZ-TH-95-30\\
November 1995\\
\end{flushright}
\vspace{0.5cm}
\begin{center}
{\Large \bf Two loop $O(N_f\alpha _s^2)$ corrections to the decay
width of the Higgs boson to two massive fermions.}
\end{center}
\vspace{1.3cm}
\begin{center}
{
K. Melnikov\footnote{Supported by Graduiertenkolleg
"Teilchenphysik", Mainz}}\\[1cm]
{\bf \em Institut f\"{u}r Physik, THEP,  Johannes
Gutenberg Universit\"{a}t,}\\
{\bf \em Staudinger Weg 7, Mainz, Germany, D 55099.}\\
\end{center}
\begin{abstract}
 We present an analytical calculation of additional real or virtual
radiation of the light fermion pair in the fermionic decay of the Higgs
boson $ H \to f_1\bar f_1$ for arbitrary ratios of the
Higgs boson mass to the $f_1$ fermion mass.
 This result gives us a value of the  $O(N_f \alpha _s^2)$
radiative correction to the inclusive decay rate $ H \to f_1\bar f_1$.
Using this result in the framework of the Brodsky-Mackenzie-Lepage scheme,
 we discuss the scale setting in the one-loop QCD correction
to the decay width $ H \to f_1\bar f_1$ for arbitrary relation between
the
Higgs boson and fermion masses.
\end{abstract}
\vspace{0.6cm}
\vspace{0.6cm}

\newpage
\section{Introduction}

Fermionic decay channels of the Standard Model (SM) Higgs boson are
important channels for both discovery and investigation of this particle
\cite {HHG}, \cite {Kniehl}.
Direct observation of such decays can give important information about
Higgs--fermion Yukawa couplings and hence provide still absent check
of the symmetry breaking mechanism in the fermion sector of the SM.

In this paper we discuss the $O(N_f\alpha _s^2)$ correction
to the decay of the Higgs boson to the pair of massive fermions $ H \to
f_1 \bar f_1$ for arbitrary relation between the Higgs mass and the mass of
the fermion $f_1$. This decay is studied in the
literature good enough. The one-loop QCD radiative correction to the fermionic
partial width
of the Higgs boson was calculated long ago \cite {Drees}. Since than, the
studies of this decay were concentrated on the analyses of the limit
 $m_H \gg m_1$ (from the phenomenological point of view this is definitely
a good approximation for the decay $H \to b\bar b$). In this limit the
renormalization group methods
were applied to this partial decay width \cite {Sakai} and the exact results
on the complete $O(\alpha _s^2)$ correction including the power suppressed
terms $O(m_1^2/m_H^2)$ were
obtained analytically \cite {Rus}, \cite {Surg}.  Recently in
the paper \cite {Chet}
these results were rederived and some new, previously missed, contributions
were calculated.

Our work is motivated by the known fact that the QCD radiative corrections
to a number of processes
involving Higgs boson - quark interactions appear to be large. It is
possible to attribute a bulk  of them to the running of the Yukawa
coupling or equivalently to the running of the fermion mass.
In this respect it is important to calculate next-to-leading order QCD
radiative corrections in order to reduce the scale ambiguity of the leading
order results and check our understanding of the resummation procedures
based on the renormalization group equations.

The complete analyses of the problem requires complete two-loop calculation of
the QCD radiative correction to the $H \to f_1 \bar f_1$ decay
channel for arbitrary relation
between Higgs and quark masses.  This task is too complicated at present.

A more easy way is provided by the Brodsky-Mackenzie-Lepage (BLM)
method \cite {BLM}. This method gives a possibility to obtain a value of
the correct
scale in the one-loop QCD correction ( and hence a good idea of a two-loop
contribution) by considering the two-loop
 diagrams, which arise due to the light fermion loop insertions into the
 gluon propagator in the one-loop QCD correction (see Fig.1).

Similar arguments and techniques were used for the analyses of the
scale setting in
the QCD radiative corrections to the $\rho$-parameter and to the
top quark decay width \cite {Vol}.

In some sense, the results presented here are complementary to the results
presented in \cite {Rus}, \cite {Surg}, \cite {Chet}
-- we calculate the part of the two-loop
QCD radiative correction, which corresponds to the running of the
 QCD coupling constant, however keeping a
relation between the Higgs boson mass and the mass of the
fermion arbitrary.  We mention here, that our
approach is similar to the one of Ref.~\cite {Teub}.

The subsequent part of the paper is organized as follows:
in the next section we discuss real radiation of the light fermion
pair in $H \to f_1 \bar f_1$; in the section 3 we analyse
virtual radiation of the fermion pair in two cases $m_1 \gg m_2$ and
   $m_1=m_2$; in the section 4 we discuss the $O(N_f\alpha _s ^2)$ correction
to the total decay width $H \to f _1 \bar f_1$; finally we present some
remarks and conclusions.

Some comments on our notations are in order. It is clear that the
major part of our discussion applies to the QED case as well. Hence, in the
first
two sections we use the QED terminology. While discussing the total decay
width $ H \to f_1 \bar f_1$  in the section 4, we switch to the QCD
notations, explicitly indicate $N_f$ dependence of the result and use
appropriate colour factors.

\section{Real decay rate $H \to f_1\bar f_1 f_2 \bar f_2$.}
\par
The Higgs boson couples to fermions proportionally to their masses. Hence,
we consider only diagrams were the Higgs boson is connected with the
heavy fermion lines.
The generic graphs are shown in the Fig.1.

\begin{figure}[htb]
\epsfxsize=12cm
\centerline{\epsffile{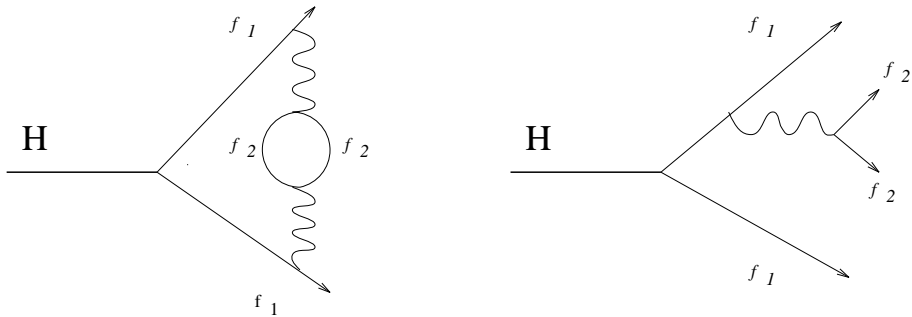}}
\caption[]{ Generic diagrams for the $O(N_f \alpha _s ^2)$ correction.
 a) Virtual radiation. b) Real radiation.}
\end{figure}

The decay rate for the process $H \to f_1 \bar f_1 f_2 \bar f_2$,
normalized to the lowest order width $\Gamma (H \to f_1 \bar f_1)$, can be
written as
 a two-dimensional integral:
$$
\frac {\Gamma (H \to f_1 \bar f_1
 f_2 \bar f_2)}{\Gamma (H \to f_1 \bar f_1)}=
 -\frac {1}{3} \Big ( \frac {\alpha}{\pi} \Big )^2 F_R
$$
$$
F_R = \frac {1}{\beta ^3}\int \limits_{4r_1}^{(1-2\sqrt{r_2})^2}dy
 \int \limits_{4r_2}^{(1-\sqrt{y})^2}
 \frac {dz}{z}(1+\frac {2r_2}{z})\sqrt{1-\frac {4r_2}{z}}\times
 \Big \{ $$
$$
\Big (\frac {-1-(y-z)^2+4r_1(1-4z+2y)-16r_1^2}{1-y+z} \Big )
\log \frac {1-y+z+\beta (y) \Lambda^{1/2}(1,y,z)}
 {1-y+z-\beta (y) \Lambda^{1/2}(1,y,z)} \nonumber $$
\begin{equation}
+\frac {4\beta (y)\Lambda^{1/2}(1,y,z)(1-4r_1)(2r_1+z)}
{(1-y+z)^2-\beta (y)^2\Lambda(1,y,z)} \Big \} \label {rate}
\end{equation}
where
\begin{equation}
 r_1=\frac {m_1^2}{m_H^2},~~~r_2=\frac {m_2^2}{m_H^2},~~~\beta =\sqrt{1-4r_1}
\end{equation}
and
\begin{equation}
\beta (y) =\sqrt{1-\frac {4r_1}{y}},~~~~~\Lambda (1,y,z)=1+y^2+z^2-2(y-z-yz)
\end{equation}
The integration variables have the following meaning: $m_H^2y$ is the
invariant mass of the pair of heavy fermions $f_1\bar f_1$ while
 $m_H^2z$ gives the invariant mass of the pair of light fermions
 $f_2 \bar f_2$.

 In the limit of interest $(m_1 \gg m_2)$ the integration can be splitted
into the ``soft'' and ``hard'' regions, depending on the energy of the
light fermion pair. Integrations over these regions are performed separately.

 Our result for the decay rate $H \to f_1\bar f_1 f_2 \bar f_2$
can be written in the following way:
\begin{equation}
F_R = f_R^{(2)}\log^2\frac {m_2^2}{m_H^2}+
 f_R^{(1)}\log\frac {m_2^2}{m_H^2} + f_R^{(0)}
\end{equation}
Below we present the functions $f_R^{(2)},f_R^{(1)},f_R^{(0)}$,
obtained by direct integration of the Eq.(\ref {rate}) \footnote{Note, that
the double logarithmic form factor is proportional to the infra-red
divergent part of the partial decay width $H \to f_1 \bar f _1 g $
 \cite {Drees}}.
\begin{equation}
 f_R^{(2)} = \frac {1+p^2}{2(1-p^2)}\log(p)+\frac {1}{2}
\end{equation}
\begin{eqnarray}
f_R^{(1)}&=& \frac {1+p^2}{1-p^2} \Big \{ 4\Li _2(-p)+6\Li _2(p)-4\zeta (2)
   + 2\log(1-p)\log(p)
 \nonumber \\
&+&4\log(1+p)\log(p)-\frac {1}{2}\log^2(p)\Big \}
-4\log(1-p)
\nonumber \\
&+&\frac {-11p^4+4p^3+20p^2-44p+13}{6(1-p)^2(1-p^2)} \log(p)
+ \frac {59(1+p^2)-136p}{12(1-p)^2}
\end{eqnarray}
\begin{eqnarray}
f_R^{(0)}&=&\frac {1+p^2}{1-p^2} \Big \{ 8\Li _3(\frac {1}{1+p})-4\Li _3(p^2)-
   8\Li _3(1-p)-10\Li _3(1-p^2)-8\Li _3(p) \nonumber \\
&+&5\zeta (3)-\frac {4}{3}\log^3(1+p)
 +\frac {1}{6}\log^3(p) -16\log(p)\log^2(1-p) \nonumber \\
&-&16\log(1+p)\log(1-p)\log(p)
+
2\log^2(p)\log(1-p) \nonumber \\
&-&12\Li _2(p)(2\log(1-p)-\log(p))-8\Li _2(-p)
(2\log(1-p)-\log(p)) \nonumber \\
&+&\zeta (2)(16\log(1-p)+4\log(1+p)+2\log(p)) \Big \} \nonumber \\
&+& \frac {4(1-3p+p^2-3p^3+p^4)}{(1-p^2)(1-p)^2}\Li _2(p)+
\frac {5+8p+28p^2+8p^3+5p^4}{3(1-p^2)(1-p)^2}\Li _2(-p) \nonumber \\
&+&\frac {17p^4+8p^3+4p^2+152p-55}{6(1-p^2)(1-p)^2} \zeta (2)
\nonumber \\
&-&\frac {2(7-26p+14p^2+22p^3-17p^4)}{3(1-p^2)(1-p)^2}\log(p)\log(1-p)
\nonumber \\
&+& 8\log(1-p)^2-\frac {59(1+p^2)-136p}{3(1-p)^2}\log(1-p)
\nonumber \\
&+&\frac {1}{36(1-p^2)(1-p)^2}\Big \{ -(575p^4-1024p^3-530p^2+608p-133)\log(p)
\nonumber \\
&-&3(13p^4+160p^3-40p^2-56p+13)\log^2(p)\nonumber \\
&+&12(5p^4+8p^3+28p^2+8p+5)\log(p)\log(1+p) \nonumber \\
&+&(1-p^2)(433(1+p^2)-830p)\Big \}
\end{eqnarray}

Here $p$ is defined through the equation:
$$ \frac {m_H^2}{m_1^2} = \frac {(1+p)^2}{p}.$$
$\Li _2 $ and $ \Li _3 $ are di--~and trilogarithms, defined in accordance with
 \cite {Lewin}.

Equations (5)-(7) give the exact result for the real decay rate
$H \to f_1 \bar f_1 f_2 \bar f_2$ in the limit $m_H  \gg m_2$, $m_1 \gg m_2$.

Let us consider now the limit of the heavy Higgs boson, which is given by
the conditions $r_1 \ll 1,~~~~r_2 \ll r_1$. In this case we expand the
complete formulae up to the terms of the order of $O(r_1)$ and get:
\begin{eqnarray}
F_R&\to &\frac {1}{2}\log(r_2)^2(\log(r_1)+1)+\log(r_2)\Big (
-\frac{1}{2}\log^2(r_1)+
\frac {13}{6} \log(r_1)-4\zeta (2)+\frac {59}{12} \Big ) \nonumber \\
&+& \frac {1}{6}\log^3(r_1)-\frac {13}{12}\log^2(r_1)+\Big (\frac {133}{36}+
 2\zeta (2) \Big ) \log(r_1)-5\zeta (3)-\frac {55}{6}\zeta(2)+\frac {433}{36}
\nonumber \\
&+&r_1 \Big [\log(r_2)^2+\log(r_2)(-3\log(r_1)+\frac {53}{6})+
\frac {3}{2}\log^2(r_1) \nonumber \\
&-&\frac {7}{2}\log(r_1)-\zeta (2)+\frac {403}{18}\Big ]
\end{eqnarray}

The opposite limit is realized when the mass of the Higgs boson is close to
two fermion masses: $r_1 \approx 1$. In this case the
velocity of the fermion is small. The photon couples to
a slow fermion  proportionally to the velocity of the latter. Hence we
expect, that in the limit $r_1 \approx 1$ the emission of the pair should
be suppressed as the square of the velocity. Calculating this limit from the
complete expression, one finds:
\begin{eqnarray}
F_R \to \beta ^2 \Big [ -\frac {2}{3} \log^2(r_2)+
 \log(r_2)\Big (\frac {16}{3}\log(2\beta)-\frac {799}{12}\Big ) \nonumber \\
-\frac {32}{3}\log^2(2\beta)+\frac {799}{18}\log(2\beta)+8\zeta (2)-
 \frac {27425}{432} \Big ]
\end{eqnarray}

\section{Virtual radiative correction.}
\subsection{General formulas.}
\par
In this section we discuss
virtual radiation of the additional fermion pair.
First we present some general formulas which are valid for arbitrary
relation between fermionic masses $m_1$ and $m_2$. Later we analyse two
cases of practical importance: $m_1 = m_2$ and $m_1 \gg m_2$.

Additional virtual radiation of the fermion pair corresponds to the
insertion of the fermion loop
to the gluon line in the one-loop QCD correction (Fig.1a).
The first step in this
consideration is to write the contribution of the light fermion pair
 to the  gluon
polarization operator through dispersion integral subtracted
at zero momentum transfer, which corresponds to the QED-like normalization
of the coupling constant \footnote{ As far as we are concerned with the
QED-like graphs such subtraction is evidently possible. Technically, it is more
convenient to change the scale of the coupling constant in the final
result, than to subtract gluon vacuum polarization at the arbitrary scale.}:
\begin{equation}
\frac {1}{k^2} \to \frac {\alpha}{3\pi}\int \limits_{4m_2^2}^{\infty}
\frac {d\lambda^2}{\lambda^2} \frac {1}{k^2-\lambda^2+i\epsilon}
\Big ( 1+\frac {2m_2^2}{\lambda^2} \Big ) \sqrt{1-\frac {4m_2^2}{\lambda^2}}
\label {polop}
\end{equation}

Due to the vector current conservation, $k_{\mu}k_{\nu}$
part of the
polarization operator does not contribute to the physical amplitude.

To evaluate the $O(N_f \alpha _s^2)$ correction to the Yukawa coupling
we consider both bare radiative correction to the triangle graph
 (Fig.1a)  and the
counterterms. In both we insert the fermion loop into the gluon line.
After writing polarization operator through dispersion integral (Eq.10),
integration over ``gluon mass'' $\lambda$ factorizes. Hence as the first step
we evaluate corrections to the Yukawa coupling coming from the
massive vector boson exchange between heavy fermions and than integrate
this result over the masses of the vector boson with the spectral density
given by the fermion contribution to the imaginary part of the gluon
polarization operator
(see Eq.(\ref {polop})). It is clear then, that we can
discuss renormalization already at the first step of our calculation.

The counterterms Lagrangian is known from the one-loop QCD
radiative correction
\cite {Drees} and can be written as \footnote{ This counterterms Lagrangian
corresponds to the on-shell subtraction of the quark mass operator. Hence
$m_1$ below is the pole mass of the quark.}:
\begin{equation}
L_{ct}=g_Y(-\Sigma_S(m,\lambda)
+2m_1^2(\Sigma_V'(m,\lambda)+\Sigma_S'(m,\lambda)))\bar \psi \psi H.
\end{equation}
Here $g_Y$ is the Yukawa coupling,
$\Sigma_{V,S}$ are defined through the quark mass operator:
\begin{equation}
\Sigma(p,\lambda) = -i\Big ( \hat p \Sigma _V(p,\lambda)
 +m_1\Sigma _S(p.\lambda) \Big )
\end{equation}
and $\Sigma _{V,S}'$ is the derivative of the corresponding quantity with
respect to $p^2$. In accordance with the preceeding discussion we explicitly
indicate a dependence of the quark mass operator on the gluon mass
$\lambda$.

To proceed further to a more precise discussion let us fix the notations.
We write the Yukawa coupling of the Higgs boson to the fermions in the
following way:
\begin{equation}
        -ig_Y T^{(2)}_V \psi \bar \psi H
\end{equation}
where $T^{(2)}_V $ is the two-loop form factor.
The sum of the bare radiative corrections to the vertex and the counterterms,
gives the following
representation for the $O(N_f\alpha _s^2)$ correction to the $Hf_1 \bar f_1$
vertex:
\begin{eqnarray}
T^{(2)}_V (\lambda^2)&=&\frac {\alpha }{3\pi}\Big \{4\pi\alpha
 \int \frac {d^4k}{(2\pi)^4}
\frac {-i(4m_1\hat k-4p_1p_2)}{(k^2-\lambda ^2)((p_1-k)^2-m_1^2)((p_2+k)^2-
 m_1^2)} \nonumber \\
&+&2m_1^2(\Sigma_V'(m_1,\lambda)+\Sigma _S'(m_1,\lambda) \Big \}
\end{eqnarray}
\begin{equation}
T^{(2)}_V = \int \limits_{4m_2^2}^{\infty}
\frac {d\lambda^2}{\lambda^2}
(1+\frac {2m_2^2}{\lambda^2})\sqrt{1-\frac {4m_2^2}{\lambda^2}}
T^{(2)}_V (\lambda^2)
\end{equation}

We assume here that the matrix element of the $\gamma $-matrices
should be evaluated with
respect to the on--shell fermion and anti-fermion spinors.

This formulae is valid for arbitrary relation between the masses $m_1$ and
$m_2$. Below we consider two special cases $m_1= m_2$ and $m_1 \gg m_2$.

In the Ref. \cite {Teub} it was suggested to calculate first the one-loop
integrals  with the arbitrary gluon mass
and than to integrate this result with the gluon spectral
density.  Here we choose a different way, which, in our opinion, is
more suitable for the two special cases we are interested in.

To demonstrate it,
we discuss below the calculation of the contribution of the scalar
three-point function (term proportional to $4p_1p_2$ in the Eq.(14)) to the
two-loop form factor $T^{(2)}_V$.

\subsection{Contribution of the scalar three-point function.}

In this subsection we indicate all the steps which are necessary to
evaluate the following integral:
\begin{equation}
I=\int \limits_{4m_2^2}^{\infty}
\frac {d\lambda^2}{\lambda^2}
\Big (1+\frac {2m_2 ^2}{\lambda^2}\Big )\sqrt{1-\frac {4m_2^2}{\lambda^2}}
C(m_H^2,\lambda^2).
\end{equation}
Here $C(m_H^2,\lambda^2)$ is the scalar three-point function:
\begin{equation}
C(m_H^2,\lambda ^2)=\int \frac {d^4k}{(2\pi)^4}
\frac {1}{(k^2-\lambda ^2)((p_1-k)^2-m_1^2)((p_2+k)^2-
 m_1^2)}
\end{equation}
Clearly, $C(m_H^2,\lambda ^2)$ is an analytic function in the complex
plane of the
$s$-variable with the cut going from $4m_1^2$ to $\infty$ along the real
axis. We write dispersion representation:
\begin{eqnarray}
C(m_H^2,\lambda ^2) = \frac {-i}{(4\pi)^2}\frac {1}{\pi}\int
\limits_{4m_1^2}^{\infty}
\frac {ds'}{s'-s-i\epsilon}C _I(s',\lambda ^2) \nonumber \\
C _I(s',\lambda ^2)= \frac {\pi}{ s' \beta(s')}
\log\Big ( \frac {s'-4m_1^2+\lambda ^2}{\lambda ^2} \Big )
\end{eqnarray}
Using this representation in the Eq.(16) and changing
the order of integration we get:
$$
I=
\frac {-i}{16\pi^2}\int \limits_{4m_1^2}^{\infty}
\frac {ds'}{(s'-s-i\epsilon)} \frac {1}{s'\beta(s')}
\int \limits_{4m_2^2}^{\infty}
\frac {d\lambda^2}{\lambda^2}
\Big (1+\frac {2m_2^2}{\lambda^2} \Big )\sqrt{1-\frac {4m_2^2}{\lambda^2}}
\log\Big ( \frac {s'-4m_1^2+\lambda ^2}{\lambda ^2} \Big )
$$

It can be seen, that in both cases of interest this representation
is very convenient for further integration.

First, in the case $m_1=m_2$
this is the one-scale integral. Integration over $\lambda ^2$ provides
a simple expression and the subsequent integration over $s'$ is
cumbersome but trivial.

The second case, $m_1 \gg m_2$ is more tricky.
If $s'-4m_1^2 \gg 4m_2^2$ the integration
over $\lambda ^2$ can be simply performed providing
a possibility to make subsequent integration over $s'$. The region of
integration,
questioning this opportunity is the
region $s'-4m_1^2 \sim 4m_2^2$. It is easy to see however, that the
contribution of this region is suppressed as $O(m_2)$. Hence it can be
completely neglected as far as we are not
interested in the light mass power corrections.

We hope that after this discussion all the steps necessary to
evaluate the virtual radiation become clear.

\subsection{Results for the virtual corrections}

Finally we present the result of our calculation of the virtual radiative
corrections. In the case {\bf $m_1 \gg m_2$ } we write the $O(N_f\alpha _s^2 )$
correction  in the form:
\begin{equation}
T^{(2)}_V =\frac {1}{6}\Big ( \frac {\alpha}{\pi} \Big )^2F^{(2)}_V
\end{equation}
and
\begin{equation}
F^{(2)}_V =
f^{(2)}_V\log^2\frac {m_2^2}{m_H^2}+
 f^{(1)}_V\log\frac {m_2^2}{m_H^2} + f^{(0)}_V
\end{equation}

The expressions for the quantities $f^{(i)}_V$ are:
\begin{equation}
 f^{(2)}_V = \frac {1+p^2}{2(1-p^2)}\log(p)+\frac {1}{2}
\end{equation}
\begin{eqnarray}
f^{(1)}_V&=& \frac {1+p^2}{1-p^2} \Big \{ -2\Li _2(p)-4\zeta (2)
   + 2\log(1+p)\log(p)
 \nonumber \\
&-&2\log(1-p)\log(p)-\frac {1}{2}\log^2(p)\Big \}
+2\log(1+p)
\nonumber \\
&+&\frac {2(1+6p+4p^2)}{3(1-p^2)} \log(p)
+ \frac {8}{3}
\end{eqnarray}
\begin{eqnarray}
f^{(0)}_V&=&\frac {1+p^2}{1-p^2}\Big \{ -2\Li _3(p)-4\Li _3(1-p)
+2\zeta (3)
+2\Li _2(p)\Big (\log(p)-2 \log(1+p)\Big ) \nonumber \\
&+&2 \zeta (2) \Big (\log(p)+6\log(1-p)-4\log(1+p)\Big )
-4\log(p)\log(1+p)\log(1-p) \nonumber \\
&+&\log^2(p)\log(1-p)
+2\log^2(1+p)\log(p)
-\log^2(p)\log(1+p) +\frac {1}{6}\log^3(p) \Big \} \nonumber \\
&+&\frac {-1}{3(1-p^2)} \Big \{ (10+24p+10p^2)(\Li _2(p)+
\log(1-p)\log(p))  \\&+&(14+48p+26p^2)\zeta (2)
+(1+6p+4p^2)(\log^2(p)
  -4\log(p)\log(1+p))\nonumber \\&-&
\frac {4}{3}(1+24p+13p^2)\log(p) \Big \}
+2\log^2(1+p)+\frac {16}{3}\log(1+p)+\frac {77}{18}
\end{eqnarray}

In the case when the Higgs boson is much heavier
than the fermion $f_1$, the expression for the $F^{(2)}_V$ reads:
\begin{eqnarray}
F^{(2)}_V&\to &\frac {1}{2}\log(r_2)^2(\log(r_1)+1)+
\log(r_2)\Big ( -\frac{1}{2}\log^2(r_1)+
\frac {2}{3} \log(r_1)-4\zeta (2)+\frac {8}{3} \Big ) \nonumber \\
&+& \frac {1}{6}\log^3(r_1)-\frac {1}{3}\log^2(r_1)+\Big (\frac {4}{9}+
 2\zeta (2) \Big ) \log(r_1)-2\zeta (3)-\frac {14}{3}\zeta(2)+\frac {77}{18}
\nonumber \\
&+&r_1 \Big [\log(r_2)^2+\log(r_2)(6\log(r_1)+\frac {4}{3})-3\log^2(r_1)
\nonumber \\
&+&16\log(r_1)-28\zeta (2)+\frac {8}{9}\Big ]
\end{eqnarray}
In the opposite case, when the Higgs boson mass is close to the threshold
of two heavy fermions, the following result can be obtained:
\begin{equation}
F^{(2)}_V \to \frac {-\pi ^2}{2\beta}\Big (
\log\Big (\frac {m_2^2}{m_H^2\beta ^2} \Big )+\frac {11}{3}\Big )+
 \log\Big ( \frac {4m_2^2}{m_H^2} \Big )-\frac {5}{6}
\end{equation}

It is instructive to combine this result with the threshold $O(\alpha)$
correction which can be found in \cite {Drees}. The result for the
threshold form factor is than:
$$
T_{V}=1+\frac {\alpha}{2\pi} \Big [ \frac {\pi^2}{2\beta} \Big (1-
\frac {\alpha}{3\pi}\Big (\log\Big (\frac {m_2^2}{m_H^2\beta ^2} \Big )+
\frac {11}{3} \Big )\Big )-1+\frac {\alpha}{3\pi} \Big (
 \log\Big ( \frac {4m_2^2}{m_H^2} \Big )-\frac {5}{6}
 \Big ) \Big ]
$$

Now it is clear, that we can eliminate large logarithms, appearing in
the next-to-leading order calculation by appropriate choice of the coupling
constant. The $\overline {MS} $ coupling constant, renormalized at the
arbitrary
scale $\mu $  can be related to the on-shell
renormalized coupling constant by the following equation:
$$ \alpha = \ams (\mu ^2)\Big (1+\frac {\ams (\mu^2)}{3\pi}
   \log\frac {m_2^2}{\mu^2} \Big )+O(\ams ^3) $$

Substituting this expression to the equation for the threshold form factor,
one finds:
\begin{eqnarray}
T_{V}&=&1+\frac {\ams (\mu^2)}{2\pi} \Big [ \frac {\pi^2}{2\beta} \Big (1-
\frac {\alpha}{3\pi}\Big (\log\Big (\frac {\mu^2}{m_H^2\beta ^2} \Big )+
\frac {11}{3} \Big )\Big ) \nonumber \\
&-&1+\frac {\alpha}{3\pi} \Big (
 \log\Big ( \frac {4\mu^2}{m_H^2} \Big )-\frac {5}{6}
 \Big ) \Big ]
\end{eqnarray}

To eliminate large logarithms appearing in this expression we have to
choose (see also \cite {Teub}) two {\it different } scales
for the $\overline {MS}$ coupling constant:
in the terms exhibiting Coulomb singularity we set $\mu^2 = m_H^2\beta^2$
 (hence the scale is given by the value of the non-relativistic three
momenta of the corresponding particles) while for the part of the
correction which does not exhibit Coulomb singularity the reasonable scale
is
$\mu ^2 =\frac {1}{4} m_H^2$.
Hence the relevant expression for the threshold form factor reads:
\begin{equation}
T_{V}=1+\frac {\pi \ams (m_H ^2\beta ^2)}{4\beta}\Big (1-
\frac {11}{9}\frac {\alpha}{\pi}\Big )-\frac{\ams (m_H^2/4)}{2\pi}
\Big (1+\frac {5}{18}\frac {\alpha}{\pi} \Big)
\end{equation}
Our discussion of the threshold region given above, is quite similar to the
discussion given in the Ref. \cite {Teub} for the threshold behaviour of the
vector current form factors (for a more detailed discussion
see Ref. \cite {BrodKuhn}). This similarity is definitely in accord with the
universality of the threshold region where dynamics is defined by a long-
range Coulomb force.

\vspace{0.5cm}
\par
Exactly the same technique as above can be used in the {\bf equal mass
case }
$m_1=m_2 $. In this case the relevant integrations can be performed very
quickly providing a simple results. The effective form factor in this
case reads:
\begin{eqnarray}
f^{(0)}_V&=&\frac {1+p^2}{1-p^2}\Big ( \frac {1}{6}\log^3(p)
-2\log(p)\zeta(2) \Big )
   +\frac {(5+4p+5p^2)(1-6p+p^2)}{6(1-p)^4}\log^2(p) \nonumber \\
&+&\frac {4(7(1+p^4)-p(44+10p+44p^2))}{9(1+p)(1-p)^3}\log(p) \nonumber \\
&-&14\zeta(2)+\frac {407(1+p^5)-1389p(1+p^3)+982p^2(1+p)}{18(1+p)(1-p)^4}
\end{eqnarray}

The threshold expansion can be obtained from the previous
equation:
\begin{equation}
f^{(0)}_V \to \frac {137}{6}-12\zeta (2) \approx 3.09412
\end{equation}

High energy expansion in this case is:
\begin{eqnarray}
F^{(2)}_V \to \frac {1}{6} \log^3(r_1)&+&\frac {5}{6} \log^2(r_1)+
  \Big ( \frac {28}{9}-2\zeta (2) \Big ) \log(r_1)\nonumber \\
  -14\zeta (2)&+&\frac {407}{18} \nonumber \\
-r_1\Big (10 \log(r_1)&+&4\zeta(2)+\frac {28}{9} \Big )
\end{eqnarray}

\section{Decay rate $H \to f_1 \bar f_1$.}

Let us now discuss the total decay rate $H \to f_1 \bar f_1$
which is obtained by summing virtual
and real corrections calculated so far~\footnote { We remind the reader,
that in this section we completely switch to the QCD terminology.
Below $\alpha _s$ always denotes the QCD
coupling constant in the $ \overline {MS}$--scheme.}.
In doing so, we find that the
double logarithms of the ratio of the square of the mass of the light
fermion to the Higgs boson mass cancel, while the single logarithmic term
survives. The coefficient of this logarithmic term is proportional
to the one-loop QCD correction to the total decay rate of $H \to f_1f_1$
\cite {Drees}~\footnote {In the Eq.(2.27) of the second reference in
 \cite {Drees} there
is a misprint. The term $
-3\log\frac {1}{1+\beta _0}\log\frac {1+\beta _0}{1-\beta _0}$ should be
written as $-3\log\frac {2}{1+\beta _0}\log\frac {1+\beta _0}{1-\beta _0}$
}. Hence, we can eliminate this large logarithm by expressing
the total decay rate $H \to f_1\bar f_1$
  through $\as (\mu ^2)$ evaluated at the scale $\mu^2=m_H^2=s$.

After that, the expression for the decay width $H \to f_1 \bar f_1$
including the $O(N_f \alpha _s^2)$ corrections reads:
\begin{equation}
\Gamma( H\to f_1 \bar f_1) = \Gamma _0\Big \{ 1+
\frac {4}{3} \Big ( \frac {\alpha _s(s)}{\pi} \Big )\delta _1+
\frac {2}{9} N_f\Big (\frac {\alpha _s(s)}{\pi} \Big )^2
\Big ( f_V^{(0)}-f_R^{(0)} \Big ) \Big \}
\end{equation}
Here $\delta _1$ is the one-loop QCD radiative correction to the
decay width (see Ref. \cite {Drees}) and $f_V^{(0)},~f_R^{(0)}$ are given by
the
Eqs.(23) and (7) respectively. $\Gamma _0$ is the Born value for the
decay width $H \to f_1 \bar f_1$:
\begin{equation}
\Gamma _0=\frac {3 G_Fm_H m_1^2}{4\pi \sqrt {2}} \beta ^3
\end{equation}

We now apply the BLM scale fixing procedure for the decay width
$H \to f_1 \bar f_1$
including the full mass dependence of the radiative corrections.
The general result for the BLM scale is than:
\begin{equation}
 \mu _{BLM} = \sqrt {s}
  \exp \Big \{ \frac {(f_V^{(0)}-f_R^{(0)})}{2\delta _1} \Big \}
\end{equation}

\begin{figure}[htb]
\epsfxsize=12cm
\centerline{\epsffile{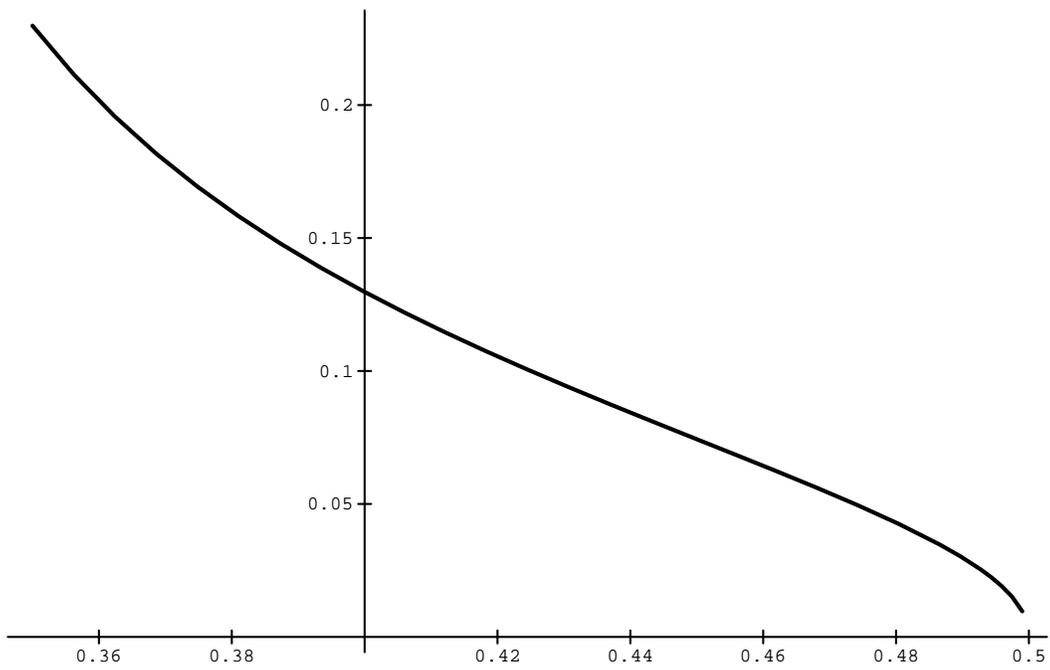}}
\caption[]{The BLM scale for the one-loop QCD correction to the
 decay width $H \to f_1 \bar f_1$
     expressed through the pole quark mass. The vertical axes is the ratio
 $\mu _{BLM}/\sqrt{s}$, the horizontal axes is the ratio
$ m/\sqrt{s}$.}
\end{figure}

\begin{figure}[htb]
\epsfxsize=12cm
\centerline{\epsffile{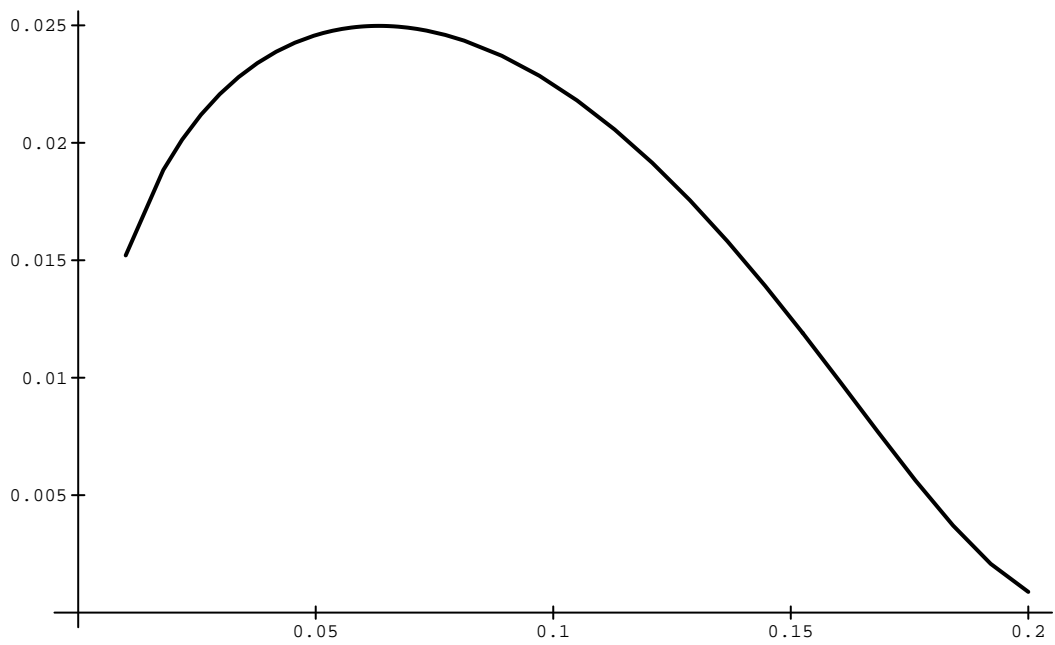}}
\caption[]{The BLM scale for the one-loop QCD correction to the
 decay width $H \to f_1 \bar f_1$
     expressed through the pole quark mass. The vertical axes is the ratio
 $\mu _{BLM}/\sqrt{s}$, the horizontal axes is the ratio
$ m/\sqrt{s}$.}
\end{figure}
The numerical results for the ratio
$\mu _{BLM} / \sqrt {s} $ as a function of the ratio
 $m_1 / \sqrt {s} $ are shown in the Figs. 2,3.
The one-loop radiative
correction
goes to zero in the vicinity of the point $m_1 / \sqrt {s} \approx 0.23$.
Around this point the BLM analyses can not be applied. The threshold BLM
scale approaches zero, in accordance with
the discussion in the section 3.  Note that the scale is quite low
even sufficiently far from the threshold.
Beyond the point
$m_1/\sqrt{s} \approx 0.23$ the BLM scale for the coupling constant
is extremely low being of the order of $\sim 0.01-0.02~m_H $.

We remind that our discussion was applied to the width expressed through
the pole mass of the quark. As was recently pointed out
 (see for instance Ref. \cite {Vol}) the low value of the BLM scale
usually encountered in such cases is connected with the fact that the pole
quark mass receives large contributions from the region of the small
loop momenta. The possible way to avoid this problem is to
express the result for the width in terms of the running quark mass.

Usually, this substitution is used in the asymptotic regime for the radiative
corrections. Here we want to check it for the whole mass range. For this
aim
we express the result for the width $H \to f_1 \bar f_1$ through
the running mass keeping only the terms of the order of
$O(N_f\alpha _s ^2)$ in the
$O(\alpha _s^2)$ corrections. The expression for the pole mass in terms of
the running mass reads:
\begin{eqnarray}
m^2 &=&\bar m ^2(\mu ^2)\Big \{ 1+\Big (\frac {\as (\mu^2)}{\pi} \Big )
\Big (-2\log\frac {\bar m ^2 (\mu^2)}{\mu ^2}+\frac {8}{3} \Big ) \\
&+&N_f\Big (\frac {\as (\mu^2)}{\pi} \Big )^2 \Big (
-\frac {1}{6}\log^2\frac {\bar m ^2 (\mu^2)}{\mu ^2}+\frac {13}{18}
\log\frac {\bar m ^2 (\mu^2)}{\mu ^2}-\frac {2}{3}\Big (
\zeta(2)+\frac {71}{48} \Big ) \Big )
\Big \} \nonumber
\end{eqnarray}
Using the expression for the width in terms of the running quark mass,
we recalculate the BLM scale. The numerical results are presented in the
Fig.4.
\begin{figure}[htb]
\epsfxsize=12cm
\centerline{\epsffile{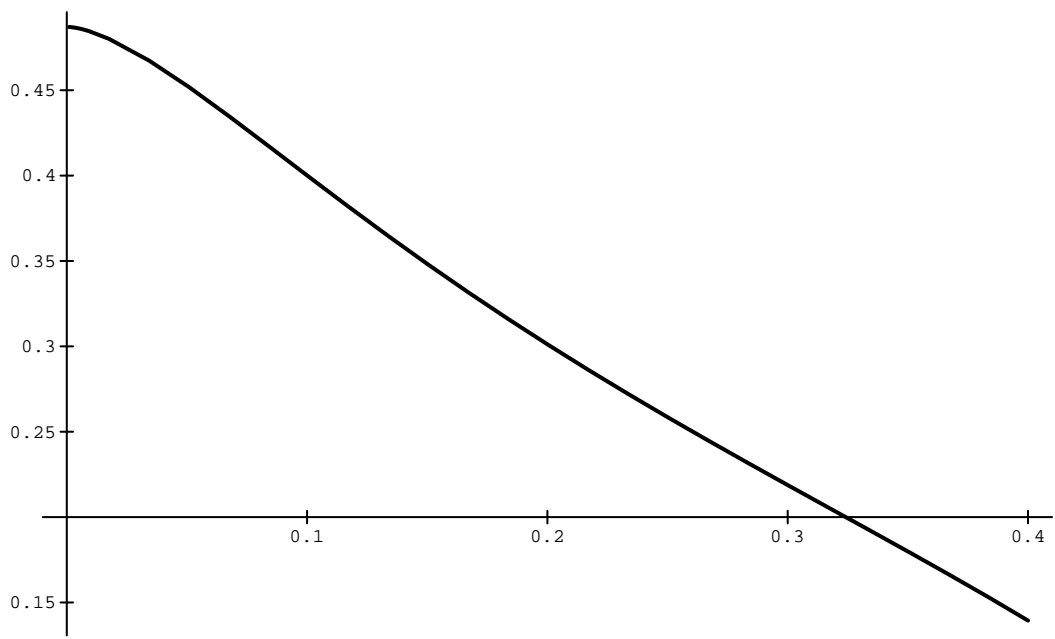}}
\caption[]{The BLM scale for the one-loop QCD correction to the
 decay width $H \to f_1 \bar f_1$
     expressed through the running mass. The vertical axes is the ratio
 $\mu _{BLM}/\sqrt{s}$, the horizontal axes is the ratio
$\bar m(s)/\sqrt{s}$.}
\end{figure}

We see that the use of the running quark mass in the expression for
the width makes the BLM scale for the coupling constant higher
for arbitrary relation between the Higgs and the fermion mass ( excluding
the region close to the threshold, where the use of the running mass is
artificial). This does not make much difference for the mass region not far
from the threshold, but for higher energies the difference is huge.
The curve on the Fig.4 can be well approximated by the
following equation:
\begin{equation}
\mu _{BLM} \approx 0.49~m_H - \bar m (m_H)
\end{equation}

The important check of our results can be performed by studying
the limit $m_H \gg m_1$. As it was mentioned above, the
$O(\alpha _s ^2)$ corrections to the decay width $ H \to f_1 \bar f_1$
are known in this limit
up to the power suppressed terms $ O(m_1^2/m_H^2)$.
Expanding the expression for the width up to the terms of the order
$O(m_1^2/s)$ and using the expression for the running mass, we get:
\begin{eqnarray}
\Gamma &=&\frac {3 G_Fm_H \bar m_1^2(s)}{4\pi \sqrt {2}}
\Big \{ 1+\frac {17}{3}
\Big (\frac {\alpha _s (s)}{\pi} \Big ) +
N_f\Big (\frac {\alpha _s(s)}{\pi} \Big )^2 \Big (\frac{2}{3}\zeta (3)
+\frac {1}{3}\zeta (2)-
\frac {65}{24} \Big ) \nonumber \\
&-&\frac {\bar m _1 ^2(s)}{s} \Big [6+40
\Big (\frac {\alpha _s(s)}{\pi} \Big )
+N_f\Big (\frac {\alpha _s(s)}{\pi} \Big )^2 \Big (4\zeta (3)+4\zeta (2)
-\frac {313}{18} \Big ) \Big ] \Big \}
\end{eqnarray}

Our result for the width in Eq.(36) is in complete agreement
with the $N_f$--dependent part of the $O(\alpha _s^2)$ correction
given in Ref. \cite {Rus}, \cite {Surg}, \cite {Chet}.

\section{Conclusion.}

By no doubts Higgs interaction with massive quarks is quite important
from phenomenological point of view. A vivid example is provided
by the decay mode $H \to \bar b b$, which can
be used for the detection of the light Higgs boson. Even leaving aside
the problem of finding this particle, direct measurement of the coupling of
the Higgs boson to quarks seems to be necessary. The measurement of such
type can test the symmetry breaking mechanism in the Higgs-fermion sector
of the Standard Model.

In this paper we present analytical results for the
$O(N_f \alpha _s^2)$ correction to the decay width of the Higgs boson
into the pair of massive fermions for the arbitrary relation between the mass
of
the Higgs boson and the mass of the fermion.
We  calculate both real and virtual
radiation of the light fermion pair in this decay. As a byproduct of this
analyses, we obtain the formulae (see Eq.(4) and below)
for the width of the rare decay:
$H \to f_1\bar f_1 f_2 \bar f_2 $ in the limit $ m_1 \gg m_2$.
\par

It seems that the most important phenomenological application of our
analyses is connected with the Higgs boson decay to two top quarks.
In this case the value of the
top quark mass and the expected value of the Higgs boson mass suggests that
there will be no small (or large) mass ratios in this problem.
In this case any results on the next-to-leading order QCD radiative
corrections are absent,
and our results provide the first step in this direction.

Our analyses of the BLM scale indicates that the use of the running quark
mass in the complete expression for the one-loop QCD radiative correction to
$H \to f_1 \bar f_1$ is definitely a good choice for arbitrary relation
between the Higgs and the fermion masses. We hope that if both
the running quark
mass and the BLM scale for the coupling constant, evaluated in
this paper (see Eq.(35)), are used for
the description of the one-loop QCD corrected decay
width $H \to f_1 \bar f_1$, this should
substitute quite reasonable approximation for the description of this
process
for arbitrary ratio of the Higgs boson mass to the fermion mass.

\end{document}